\documentclass[twocolumn,showpacs,pre,preprintnumbers,amsmath,amssymb]{revtex4-2}

\usepackage{graphicx}
\usepackage{dcolumn}
\usepackage{bm}
\usepackage{appendix}
\usepackage{hyperref}
\usepackage[capitalise]{cleveref}
\usepackage{booktabs} 
\usepackage{multirow} 
\usepackage{subcaption} 
\usepackage{ragged2e}
\DeclareCaptionJustification{justified}{\justifying}
\usepackage{xspace}

\def\CeRuFeGe{Ce(Fe$_{0.76}$Ru$_{0.24}$)$_2$Ge$_2$\xspace}

\begin{document}

\title{The effects of disorder on Harris-criterion violating percolation}
\author{Sean Fayfar, Alex Breta\~{n}a, and Wouter Montfrooij}
\affiliation{Department of Physics and Astronomy, University of Missouri, Columbia, MO 65211, United States}

\begin{abstract}
    We present the results of computer simulations on a class of percolative systems, called protected percolation, that violates the Harris criterion. The Harris criterion states whether the critical behavior at a phase transition from a disordered state to an ordered state will be altered by impurities. We have incorporated impurities into our simulations to test whether the critical exponents for protected percolation are altered by impurities. We find that the critical exponents for three-dimensional protected percolation simulations indeed change with impurities in the form of missing sites and immortal sites. On the other hand, the critical exponents for both standard percolation and protected percolation in two dimensions are stable against impurities. 
\end{abstract}

\date{\today}

\maketitle

\section{Introduction}\label{sec:introduction}
Percolation theory is the study of the transitions that occur upon the breaking of connections in a system through the removal of elements or the connections between them. Percolation applies to many physical systems and falls into known universality classes \cite{Stauffer1985,Nakayama1994,Sahini1994}. When enough of the connections have been broken, the system spanning connection will be fractured at a point defined as the percolation threshold; the point at which this happens is dependent on both the dimensionality and the connectivity of the system. However, when approaching this threshold, universal behavior is displayed that only depends on the dimensionality. 

Protected percolation -- described in detail in a previous paper \cite{Fayfar2021} -- has the added restriction that upon emptying a lattice, only sites from the system spanning cluster are removed. Because of this restriction, the isolated clusters that form will be protected from any further removal of their sites. As such, they will have a unique morphology compared to the isolated clusters in standard percolation. The lattice spanning cluster will undergo the same random removal process in both standard and protected percolation. Since only sites from the lattice spanning cluster can be removed in protected percolation, the strength of the lattice spanning cluster will diminish more rapidly. Protected percolation forms its own universality class abs has its own set of critical exponents that relate to those of standard percolation through analytical relationships \cite{Fayfar2021}.

The Harris criterion predicts whether the universal behavior of a transition from a disordered state to an ordered state will be stable against impurities \cite{Harris1974}. The original derivation considered both a randomly diluted system as well as a system of spins; we will discuss the latter. When we have a spin system with small random variations in the interaction strengths, then the system will fracture into subvolumes, each with a slight variation in transition temperature. This spread in interaction strength creates a concomitant spread in transition temperatures within the system. The spread in transition temperatures will not be relevant to the critical behavior if the spread in temperatures goes to zero faster than the system's approach to the transition temperature \cite{Harris1974}. The criterion is written mathematically using the critical exponent for the correlation length $\nu$ and the dimensionality $d$ \cite{Harris1974}:
\begin{equation}
    \label{eq:harrisCriterion}
    d \nu = \gamma + 2\beta > 2,
\end{equation}
where scaling laws relate the critical exponents for the order parameter $\beta$ and the average cluster size $\gamma$ \cite{Stauffer1985} to $\nu$. 

Protected percolation violates the Harris criterion in three dimensions and satisfies it in two dimensions \cite{Fayfar2021}. In contrast, standard percolation satisfies it in both cases \cite{Fayfar2021}. As such, any system that obeys a three-dimensional protected percolation model should have unique critical exponents dependent on the inherent impurities in each system; our aim for this paper is to investigate that claim and test what effects impurities have on protected percolation. Protected percolation simulations provide a perfect platform to test the Harris criterion and the significance of its violation. Only a few known systems violate the Harris criterion and provide a computationally testable platform \cite{Jensen1996,Sknepnek2004,Yao2010,Birgeneau1983}. 

The outline of this paper is as follows. In the remainder of this section, we review the relevant equations for percolation. In \cref{sec:methods} we describe the types of impurities and how they are implemented in our simulations. In \cref{sec:results} we give the results for the critical exponents with impurities. We discuss our findings in \cref{sec:discussion} and detail the connections to physical systems. 

In site percolation \cite{Stauffer1985}, the probability that a random site is occupied is defined as the occupation $p$; the point when the lattice spanning connection fractures is defined as the percolation threshold and is denoted by $p_c$. The $k^\textrm{th}$ moment of the cluster size distribution is defined as
\begin{equation}
    \label{eq:clusMoments}
    M_k(p) \equiv \sum_s s^k n_s(p),
\end{equation}
where $s$ is the number of sites in the cluster, and $n_s$ is the number of clusters per site containing $s$ sites \cite{Stauffer1985}. The strength of the lattice spanning cluster $P(p)$ is the number of sites in the lattice spanning cluster divided by the number of lattice sites and is related to the $1^\textrm{st}$ moment through
\begin{equation}
    \label{eq:strPercClus}
    P(p) = p - \sum_s s n_s(p).
\end{equation}
The average cluster size $S(p)$ is defined as 
\begin{equation}
    \label{eq:clusSize}
    S(p) \equiv \frac{\sum_s s^2 n_s(p)}{\sum_s s n_s(p)} \approx \frac{\sum_s s^2 n_s(p)}{p_c},
\end{equation}
where it is approximately equivalent to the second moment ($k=2$) close to the percolation threshold \cite{Stauffer1985}. 

When approaching the percolation threshold, critical behavior manifests and it can be modeled with power law exponents -- called critical exponents -- that only depend on the dimensionality of the system. The strength of the lattice spanning cluster displays non-analytical critical behavior, whereas the average cluster size diverges. They are modeled by the following power laws
\begin{align}
    P(p) &= P_0 (p-p_c)^\beta + (p-p_c)
    \label{eq:strpercclusPowerLaw}\\
    S(p) &= S_0 (p-p_c)^{-\gamma}
    \label{eq:clussizePowerLaw}
\end{align}
where $\beta$ and $\gamma$ are the aforementioned critical exponents \cite{Stauffer1985}. 

\section{Methods}\label{sec:methods}
We use Monte Carlo simulations to determine the critical exponents for both standard and protected percolation. We use a disjoint-set data structure \cite{Galler1964} developed by \citet{Newman2000,Newman2001} to analyze the connectivity of the clusters. The algorithm starts with an empty lattice and adds sites one at a time until it is filled. Since protected percolation allows sites to be removed from the lattice spanning cluster only, we need to retroactively correct for the sites that should not have been added into the lattice under protected percolation. An example of this would be the merging of two clusters upon filling, which would correspond to removing a site from an isolated cluster upon emptying. Using retroactive corrections has the benefit that we simulate both standard and protected percolation simultaneously. The details of the procedure can be found in \cite{Fayfar2021}.

To test the predictions of the Harris criterion \cite{Harris1974}, we added impurities into our simulation in the form of an imperfect lattice structure, missing sites, and immortal sites. Typically, we define our lattice structure through the use of a connectivity function that determines the locations of the neighbors for each site. To create an imperfect structure, we adjusted the connectivity function to include more or fewer neighboring sites than average. For example, a simple cubic structure will have six neighbors for each site, and we chose to remove some of those sites at random for a fraction of the lattice. Missing sites are sites that, upon filling, should be added into our lattice but never actually make it into the lattice. Similarly, immortal sites are sites that, upon emptying, should be removed but never actually get removed from the lattice. The latter two of these are impurities in the occupation $p$ of the lattice. An example for immortal sites would be if we have our occupation at $p=0.50$, we might have an actual lattice occupation of $0.55$ with $5\%$ of those sites non-removable. Note that such an impurity invalidates \cref{eq:strPercClus}. 

It is straightforward to add impurities into simulations. First, the incidence of these impurities must be chosen. Then during the simulation, we generate a double-precision floating-point random number and add an impurity at that site if the random number is less than the incidence amount. To create an imperfect lattice structure, we determine the neighbors that our site should have and either remove some of them or include additional next-nearest-neighbors; this gives our lattice slightly shorter or longer range connectivity. For missing sites, we skip over the addition of a site and still increment the occupancy. Immortal sites will take a few more steps since we are filling our lattice. The locations of the immortal sites need to be randomly chosen and added into the simulation at the start without incrementing the occupancy. We then fill the lattice normally, and when an immortal site occurs during the simulation process, we must skip over them since they have already been added, which would correspond to being unable to remove them upon emptying. 

\begin{figure}[t]
    \includegraphics[width=\columnwidth]{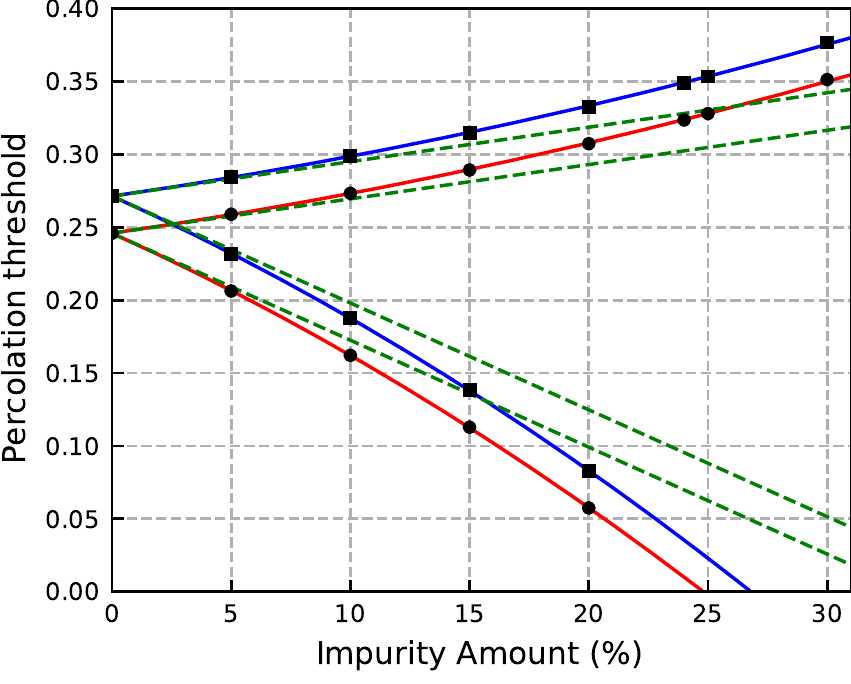}
    \caption{\label{fig:impurityThreshold}Simulation results for the percolation thresholds for standard (circle) and protected (square) percolation with impurities for a body-centered structure. Missing sites and immortal sites are shown as the top and bottom curves, respectively. The linear terms for all data sets are shown as dashed lines; the solid lines are quadratic fits.}
\end{figure}

\section{Results}\label{sec:results}
\begin{figure*}[t]
    \begin{minipage}[b]{0.49\linewidth}
        \includegraphics[]{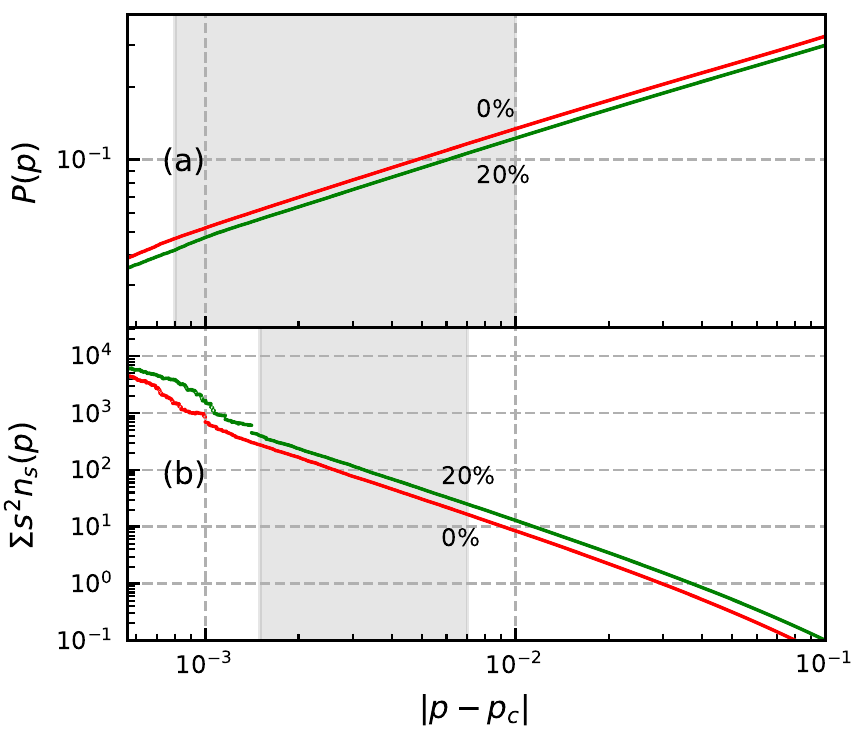}
        \phantomsubcaption\label{fig:strPercStd}
        \phantomsubcaption\label{fig:clusSizeStd}
    \end{minipage}
    \begin{minipage}[b]{0.5\linewidth}
        \includegraphics[]{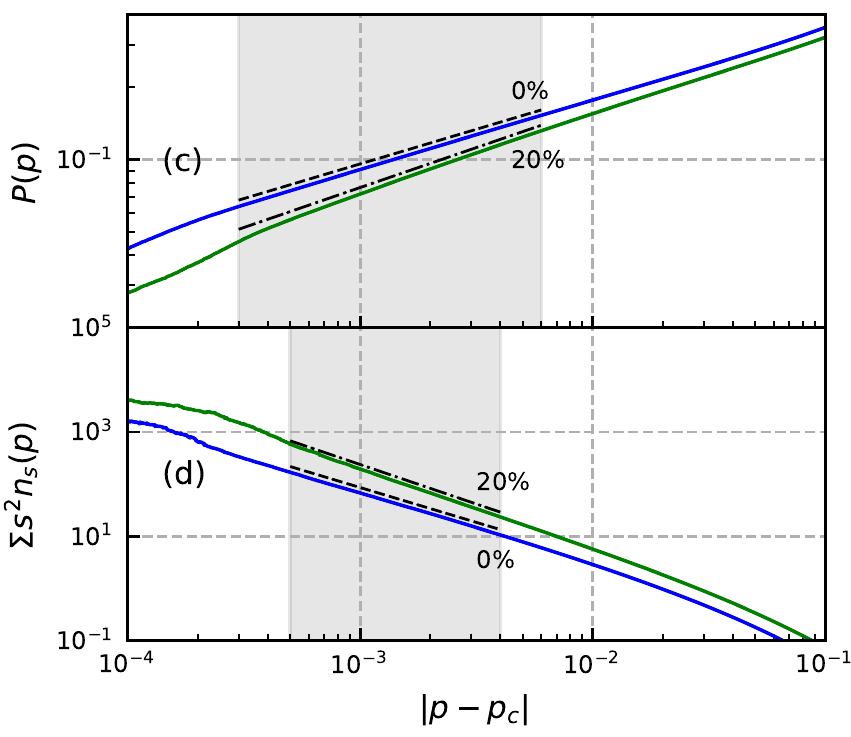}
        \phantomsubcaption\label{fig:strPercPro}
        \phantomsubcaption\label{fig:clusSizePro}
    \end{minipage}

    \caption{\label{fig:impurityCritExp}The results of simulations of size $400^3$ averaged over 1000 iterations for a body-centered structure of a simulation without impurities compared to the addition of 20\% immortal sites. (panels a,c) Log-log plot of the strength of the lattice spanning cluster -- see \cref{eq:strPercClus,eq:strpercclusPowerLaw}. (panels b,d) Log-log plot of the second moment of the cluster size distribution -- see \cref{eq:clusSize,eq:clussizePowerLaw}. The left panels are for standard percolation, and the right panels are for protected percolation; the value of the critical exponents changes with the addition of impurities for protected percolation (non-parallel lines), but not in standard percolation (parallel lines) -- see \cref{tab:critExpImp}. The gray region highlights the critical region where we determine our exponents through a fitting procedure.}
\end{figure*}

We completed computer simulations with an imperfect lattice structure that had $20\%$ of the sites with fewer neighbors than expected; we found that this shifted the percolation threshold but did not change the critical exponents. If this type of impurity were to affect the critical exponents, it should have done so given the amount. This result was expected because, as a universality class, the critical exponents (should) only depend on the dimensionality and not on the details within the structure, such as the number of neighbors. Adjusting the connectivity range skews the lattice, but the phase transition displays the same critical exponents. Given this, we do not list the values for the percolation thresholds for this particular impurity since we are interested in impurities that actually affect the critical exponents for our test of the Harris criterion. Nonetheless, this result underscores the validity of our computational approach.

Impurities, in the form of missing sites and immortal sites, do affect the critical exponents for protected percolation but not for standard percolation; the remainder of this section focuses on these types of impurities. The percolation threshold shifts with the addition of these impurities, and \cref{fig:impurityThreshold} shows that the shift is quadratic in the impurity amount. We determine the percolation threshold by using the Levinshtein method \cite{Levinshtein1975}; this entails determining the percolation threshold for 1000 iterations as a function of the lattice size. The distribution of thresholds becomes more narrow as the lattice size increases, and we extrapolate to a delta-function distribution representative of an infinite lattice \cite{Levinshtein1975,Fayfar2021}.

The critical exponents change with the addition of impurities in three-dimensional protected percolation, but not in standard percolation -- see \cref{tab:critExpImp,fig:impurityCritExp}. We completed simulations of size $400^3$ averaged over 1000 iterations using a body-centered lattice structure with the addition of missing and immortal sites. The critical exponents for standard percolation vary slightly but remain essentially unchanged within the increased uncertainty estimate. The fact that we see the critical exponents hover around a constant value as opposed to continually increasing indicates that they are stable against the impurities. The deviations for protected percolation increase with the addition of impurities in the form of immortal and missing sites, even when considering the uncertainty estimate in extracting the critical exponents. This result confirms that for standard percolation satisfying the Harris criterion, the critical exponents do not change with impurities. However, protected percolation violates the criterion, and the critical exponents do change with impurities.

\begin{table*}[th]
    \caption{\label{tab:critExpImp}The critical exponents for both standard and protected percolation with the addition of impurities in the form of missing and immortal sites. These exponents were determined by fitting results from computer simulations of size $400^3$ for 3D and $8000^2$ for 2D. All simulations were averaged over 1000 iterations with a body-centered and square structure, respectively. The error bars listed in the table were determined by analyzing the variation in the critical exponents from the uncertainty in the percolation thresholds. The missing entries are where the fit range became too constricted for accurate determination of the exponents.}
    \begin{tabular*}{\textwidth}{l @{\extracolsep{\fill}} llllllll} \toprule
        & \multicolumn{2}{c}{Impurity} & \multicolumn{3}{c}{Standard Percolation} & \multicolumn{3}{c}{Protected Percolation} \\
        \cmidrule(lr){2-3} \cmidrule(lr){4-6} \cmidrule(lr){7-9}
        Dimensions & Type & Amount & $\beta$ &  $\gamma$ & $\gamma + 2 \beta$ & $\beta'$ & $\gamma'$ & $\gamma'+2\beta'$ \\ \midrule
        \multirow{6.5}{*}{3D} & None & 0\% & 0.4053(5) & 1.819(3) & 2.6296(32) & 0.28871(15) & 1.3066(19) & 1.8840(19)\\
        \cmidrule(){2-9}
        & \multirow{6}{*}{Missing} & 2\% & 0.403(2) & 1.81(1) & 2.62(1) & 0.292(1) & 1.35(7) & 1.93(7)\\
        & & 10\% & 0.405(5) & 1.80(3) & 2.61(3) & 0.312(3) & 1.44(2) & 2.06(2)\\
        & & 20\% & 0.407(7) & 1.81(4) & 2.62(4) & 0.331(7) & 1.52(4) & 2.18(4)\\
        & & 30\% & 0.408(9)	 & 1.81(6) & 2.63(6) & 0.35(1) & 1.61(7) & 2.32(7)\\
        & & 50\% & 0.41(1) & 1.82(8) & 2.64(8) & 0.38(2) & 1.73(12) & 2.5(1)\\
        & & 70\% & 0.41(6) & - & - & 0.40(7) & - & - \\
        \cmidrule(){2-9}
        & \multirow{2}{*}{Immortal} & 10\% & 0.4079(12) & 1.830(7) & 2.646(7) & 0.3116(5) & 1.442(4) & 2.065(4)\\
        & & 20\% & 0.4058(8) & 1.795(5) & 2.607(5) & 0.3325(5) & 1.515(3) & 2.180(3)\\
        \midrule
        \multirow{2.5}{*}{2D} & None & 0\% & 0.1386(5) & 2.39(2) & 2.67(1) & 0.1219(4) & 2.105(6) & 2.349(6) \\
        \cmidrule(){2-9}
        & Missing & 20\% & 0.136(3) & 2.34(8) & 2.61(8) & 0.124(4) & 2.14(7) & 2.39(7)\\
        \bottomrule
    \end{tabular*}
\end{table*}

We also verified that the critical exponents are stable against impurities in two-dimensional standard and protected percolation -- see \cref{tab:critExpImp}. We completed simulations of size $8000^2$ averaged over 1000 iterations for a square lattice structure with the addition of missing sites. The critical exponents did not vary significantly compared to the exponents without impurities. Even though the estimated uncertainty increased with impurity concentration, we could still conclude that the exponents for simulations with impurities overlapped with the values without impurities. Given that the impurity amount -- 20\% in this case -- should have been significant enough to induce a change, we conclude that two-dimensional standard and protected percolation are stable against impurities. 

The value of the critical exponents in \cref{tab:critExpImp} for protected percolation increases with the percentage of impurities to the point where the Harris criterion becomes satisfied ($\gamma + 2\beta >2$). Interestingly, they continue to increase with an increase of impurities even when the exponents already satisfy the criterion. Similar to our findings, \citet{Jensen1996} also found that for directed percolation, the addition of disorder leads to critical exponents that change continuously with the strength of the disorder \cite{Broadbent1957}. 

\section{Discussion}\label{sec:discussion}
Protected percolation was designed to model the cluster formation in heavily-doped quantum critical compounds such as \CeRuFeGe \cite{Montfrooij2007}; here, we will briefly summarize the details -- a more in-depth discussion is given in \cite{Fayfar2021}. In systems such as \CeRuFeGe, Kondo shielding effectively removes magnetic moments from the lattice upon cooling, and the temperature at which magnetic moments are removed depends sensitively on the interatomic distances. The Fe/Ru doping creates a distribution of interatomic distances and thereby creates a distribution of Kondo shielding temperatures. Thus, each of the Ce-ions will be Kondo shielded at a unique temperature, and a percolation network will ensue upon cooling. Once clusters peel off the system spanning connection, they will align with their neighbors because of quantum mechanical finite-size effects \cite{Montfrooij2019,Heitmann2014}. The isolated clusters are protected from further moment removal as the Kondo screening mechanism involves a spin-flip process \cite{Sachdev2011,Kondo1964} that is severely impeded in an ordered environment. Thus, a protected percolation network will form upon cooling.

The immortal site impurities mimic the impurities found in heavily-doped quantum critical systems. Upon cooling heavily-doped quantum critical compounds, some magnetic moments will not be shielded since their local surroundings are such that their Kondo temperature is extremely low, with the result that they remain present in the magnetic lattice down to the lowest temperatures. Universal critical exponents have not been found for quantum critical compounds despite being widely studied \cite{Stewart2001,*Stewart2006}. Our results suggest that universal critical exponents do not exist for systems that follow protected percolation due to the impurities associated with doping changing the critical exponents. Of course, critical exponents in quantum critical systems are determined as a function of temperature, not as a function of occupancy. When the underlying critical behavior is regulated by occupancy, then different critical systems will display different critical behavior when their occupancy as a function of temperature, $p(T)$, differs. However, our findings go one step further: even when it is possible to disentangle $p(T)$ from the critical behavior as a function of temperature, then we still predict that real differences remain, reflecting the violation of the Harris criterion. 

We find that the critical exponents change in three-dimensional protected percolation but not in two-dimensions as is predicted by the Harris criterion \cite{Harris1974}. It is not clear to us why impurities play a larger role in three dimensions than in two dimensions. For stability, the Harris criterion requires that the width of the distribution of transition temperatures of the collection of sub-volumes must shrink faster than the rate at which a system approaches the transition temperature \cite{Harris1974}. Apparently, this condition remains satisfied in protected percolation in two dimensions but not in three dimensions. Perhaps the fact that in protected percolation, the shift in threshold (compared to standard percolation) increases with an increased number of dimensions \cite{Fayfar2021} might be relevant to this observation, but we cannot be sure at present. 

The critical exponents for protected percolation approach those of standard percolation with the addition of impurities. This is likely because of the nature of the impurities that we have implemented. For immortal sites emptying a lattice, we attempt to remove sites from the lattice, but some sites cannot be removed, including sites in the lattice spanning cluster. Normally in protected percolation, the strength of the lattice spanning cluster will diminish at every site removal. Since some of those sites cannot be removed with immortal-site impurities, the lattice spanning cluster does not decrease nearly as quickly as a function of decreasing occupancy, and hence, the critical exponents for protected percolation edge closer to those of standard percolation. 

We find that the critical exponents change continuously with the addition of impurities in three-dimensional protected percolation -- a result similar to that of \citet{Jensen1996}. \Citet{Jensen1996} tested the impact of temporal disorder on directed percolation and found that the critical exponents change continuously with the strength of the disorder. Note, the directed percolation critical exponents not only change continuously but continue to change after the Harris criterion is satisfied. \Citet{Chayes1986} predicted that in cases where the Harris criterion is violated, a new critical point arises with critical exponents that satisfy the criterion. Our results verify that predictions made by \citet{Chayes1986}. These results suggest that only the critical exponents without disorder predict whether a phase transition is stable against impurities. 

In conclusion, the critical exponents for protected percolation in three dimensions change continuously with the addition of impurities in the form of missing and immortal sites. The critical exponents for protected percolation in two-dimensions and standard percolation are stable against these impurities, exactly as predicted by the Harris criterion \cite{Harris1974}. The critical exponents continue to change even after they satisfy the Harris criterion suggesting that only the critical exponents without (or small amounts of) impurities predict stability. 

\bibliography{Percolation}

\end{document}